\documentclass[aps,pra,twocolumn,groupedaddress,floatfix]{revtex4}%
\usepackage{graphicx}
\usepackage{dcolumn}
\usepackage{bm}
\usepackage{hyperref}
\usepackage{amssymb}
\usepackage{amsmath}
\usepackage{amsfonts}
\usepackage{amssymb}
\usepackage{color}%
\begin{document}
\title{Controlled state-to-state atom-exchange reaction in an ultracold atom-dimer mixture}
\author{Jun Rui}
\thanks{These authors contributed equally to this work.}
\affiliation{Shanghai Branch, National Laboratory for Physical Sciences at Microscale and
Department of Modern Physics, University of Science and Technology of China,
Hefei, Anhui 230026, China}
\affiliation{CAS Center for Excellence and Synergetic Innovation Center in Quantum
Information and Quantum Physics, University of Science and Technology of
China, Shanghai 201315, China}
\affiliation{CAS-Alibaba Quantum Computing Laboratory, Shanghai 201315, China}
\author{Huan Yang}
\thanks{These authors contributed equally to this work.}
\affiliation{Shanghai Branch, National Laboratory for Physical Sciences at Microscale and
Department of Modern Physics, University of Science and Technology of China,
Hefei, Anhui 230026, China}
\affiliation{CAS Center for Excellence and Synergetic Innovation Center in Quantum
Information and Quantum Physics, University of Science and Technology of
China, Shanghai 201315, China}
\affiliation{CAS-Alibaba Quantum Computing Laboratory, Shanghai 201315, China}
\author{Lan Liu}
\affiliation{Shanghai Branch, National Laboratory for Physical Sciences at Microscale and
Department of Modern Physics, University of Science and Technology of China,
Hefei, Anhui 230026, China}
\affiliation{CAS Center for Excellence and Synergetic Innovation Center in Quantum
Information and Quantum Physics, University of Science and Technology of
China, Shanghai 201315, China}
\affiliation{CAS-Alibaba Quantum Computing Laboratory, Shanghai 201315, China}
\author{De-Chao Zhang}
\affiliation{Shanghai Branch, National Laboratory for Physical Sciences at Microscale and
Department of Modern Physics, University of Science and Technology of China,
Hefei, Anhui 230026, China}
\affiliation{CAS Center for Excellence and Synergetic Innovation Center in Quantum
Information and Quantum Physics, University of Science and Technology of
China, Shanghai 201315, China}
\affiliation{CAS-Alibaba Quantum Computing Laboratory, Shanghai 201315, China}
\author{Ya-Xiong Liu}
\affiliation{Shanghai Branch, National Laboratory for Physical Sciences at Microscale and
Department of Modern Physics, University of Science and Technology of China,
Hefei, Anhui 230026, China}
\affiliation{CAS Center for Excellence and Synergetic Innovation Center in Quantum
Information and Quantum Physics, University of Science and Technology of
China, Shanghai 201315, China}
\affiliation{CAS-Alibaba Quantum Computing Laboratory, Shanghai 201315, China}
\author{Jue Nan}
\affiliation{Shanghai Branch, National Laboratory for Physical Sciences at Microscale and
Department of Modern Physics, University of Science and Technology of China,
Hefei, Anhui 230026, China}
\affiliation{CAS Center for Excellence and Synergetic Innovation Center in Quantum
Information and Quantum Physics, University of Science and Technology of
China, Shanghai 201315, China}
\affiliation{CAS-Alibaba Quantum Computing Laboratory, Shanghai 201315, China}
\author{Bo Zhao$^{\dagger}$}
\thanks{bozhao@ustc.edu.cn}
\affiliation{Shanghai Branch, National Laboratory for Physical Sciences at Microscale and
Department of Modern Physics, University of Science and Technology of China,
Hefei, Anhui 230026, China}
\affiliation{CAS Center for Excellence and Synergetic Innovation Center in Quantum
Information and Quantum Physics, University of Science and Technology of
China, Shanghai 201315, China}
\affiliation{CAS-Alibaba Quantum Computing Laboratory, Shanghai 201315, China}
\author{Jian-Wei Pan$^{\dagger}$}
\thanks{pan@ustc.edu.cn}
\affiliation{Shanghai Branch, National Laboratory for Physical Sciences at Microscale and
Department of Modern Physics, University of Science and Technology of China,
Hefei, Anhui 230026, China}
\affiliation{CAS Center for Excellence and Synergetic Innovation Center in Quantum
Information and Quantum Physics, University of Science and Technology of
China, Shanghai 201315, China}
\affiliation{CAS-Alibaba Quantum Computing Laboratory, Shanghai 201315, China}

\maketitle

\textbf{Ultracold molecules offer remarkable opportunities to study chemical
reactions at nearly zero temperature \cite{Krems2008, Carr2009, Bell2009,
Stwalley2004, Quemener2012}. Although significant progresses have
been achieved in exploring ultracold bimolecular reactions
\cite{Ospelkaus2010a, Ni2010,Zahzam2006, Staanum2006, Hudson2008,Wang2013},
the investigations are usually limited to measurements of the overall loss rates of
the reactants. Detection of the reaction products will shed new light on
understanding the reaction mechanism and provide a unique opportunity to study
the state-to-state reaction dynamics \cite{Upadhyay2006,
Knoop2010}. Here we report on the direct observation
of an exoergic atom-exchange reaction in an ultracold atom-dimer mixture. Both
the atom and molecule products are observed and the quantum states are
characterized. By changing the magnetic field, the reaction can be switched on
or off, and the reaction rate can be controlled. The reaction is efficient and
we have measured a state-to-state reaction rate of up to}
\textbf{$1.1(3)\times10^{-9}$~cm$^{3}/$s from the time evolution of
the reactants and products. Our work represents the realization of a controlled quantum state
selected/resolved ultracold reaction. The atom-exchange reaction observed is
also an effective spin-exchange interaction between the Feshbach molecules and
the fermionic atoms and may be exploited to implement quantum simulations
of the Kondo effect with ultracold atoms and molecules \cite{Bauer2013}.}

\vspace{0.5cm}

The atom-exchange reaction we study is a bimolecular reaction,
${AB+C\rightarrow AC+B}$, where ${AB}$ and ${AC}$ are both weakly bound
Feshbach molecules, and $A$, $B$ and $C$ are distinguishable atoms. As the
binding energy of Feshbach molecules can be precisely controlled by changing
the external magnetic field \cite{chin2010}, it is possible to tune the energy
released in the reaction to be small so that the reaction products can still be
trapped. In this case, the quantum states of the reaction products may be
detected and characterized, and the state-to-state reaction dynamics may be studied.

Scientific interest in the reactive collision involving the uppermost
long-range molecules dates back about 40 years and has revived recently due to
the creation of Feshbach molecules \cite{Stwalley1978,Stwalley2004}. It was
speculated by W. Stwalley that this kind of reactive collisions for long-range
molecules may be qualitatively different from collisions involving only normal
or vibrationally excited molecules, and the reaction might selectively react
through a single channel ${AB+C\rightarrow AC+B}$ even if many inelastic
collision and reaction channels are energetically allowed. Investigations of
such atom-exchange reactions can only be conducted in ultracold gases so far
since the near dissociation molecules can be easily destroyed by thermal collisions.

Earlier experimental studies of this kind of reaction ${A}_{2}+B\rightarrow
AB+A$ were performed in ultracold Cs gases with $A$ and $B$ being different internal
states \cite{Knoop2010}. The observation of the atom product was reported,
while the molecule product ${AB}$ could not be observed because of the absence of
Feshbach resonances and the short lifetime of the ${AB}$ molecules. Besides that, only
the loss rates of the molecule reactants were measured. Another experiment that
is related to the exchange reaction was performed with three internal states
of $^{6}\mathrm{Li}$, where the influence of the reaction on the overall loss rates
was observed \cite{Lompe2010,Nakajima2010}. However, the observation of the
reaction products was impossible in this work, as the reaction is strongly suppressed
for magnetic fields at which the reaction products can be trapped. Therefore, a
definitive detection of this kind of atom-exchange reaction and study of the
reaction dynamics still remains elusive.

Here we report on the direct observation of the ${AB+C\rightarrow AC+B}$
reaction and the study of the reaction dynamics in an ultracold $^{23}$Na and
$^{40}$K mixture, where $A$ is the lowest ground hyperfine state
$|F,m_{F}\rangle_{\text{Na}}=|1,1\rangle$ of $^{23}$Na, and $B$ and $C$ are the
$|F,m_{F}\rangle_{\text{K}}=|9/2,-5/2\rangle$ and $|9/2,-3/2\rangle$ ground
hyperfine states of $^{40}$K, respectively. The ${AB}$ and ${AC}$ molecules
are the corresponding NaK Feshbach molecules denoted by $|1,1;9/2,-5/2\rangle$
and $|1,1;9/2,-3/2\rangle$. Since $B$ and $C$ are different internal states of
the same atom, the atom-exchange reaction can be regarded as an effective
spin-exchange interaction between an NaK Feshbach molecule and a K atom. In
our experiment, we prepare approximately $3.0\times10^{5}$ Na and
$1.6\times10^{5}$ K atoms at about $500$ nK in a crossed-beam dipole trap (see
Methods). The measured trap frequencies for K are $h\times(235,221,83.5)$ Hz and
the trap depth is calculated to be about $k_{\mathrm{B}}\times5$ $\mu$K for K.

\begin{figure}[ptb]
\includegraphics[width=7cm]{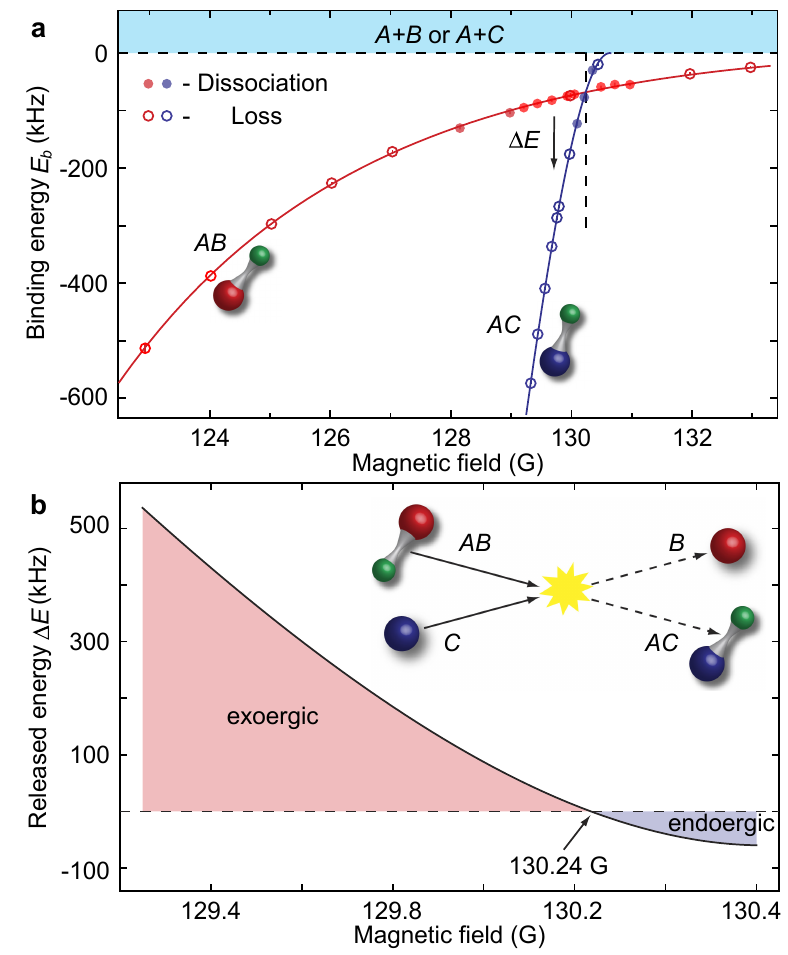}\caption{\textbf{Binding energy of the
Feshbach molecules and released energy in the reaction.} \textbf{a,}
The binding energies of the $AB$ and $AC$ molecules are measured via rf
dissociation \cite{Wu2012,Chin2005} or loss spectra \cite{Ulmanis2015}.
The data is fitted using the universal model \cite{chin2010}. \textbf{b,}
Released energy in the $AB+C\rightarrow AC+B$ atom-exchange
reaction as a function of the magnetic field. The released energy of the
reaction is given by the difference between the binding energies of the
$AB$ and $AC$ molecules. At 130.24 G, the $AB$ and $AC$ molecules
have almost the same binding energy. Below this field, the reaction is
exoergic, while above this field, the reaction is endoergic. }%
\label{fig1}%
\end{figure}

We first characterize the Feshbach resonance between the $A$ and $B$ ($C$) atoms
at about 138 G (131 G) \cite{Park2012} by measuring the binding energies of
the $AB$ ($AC$) molecules. The results are shown in Fig. \ref{fig1}\textbf{a}
and the data is fitted using the universal model \cite{chin2010}, which gives
resonance positions of $B_{0}=138.71(20)$ G and $130.637(14)$ G for the two
resonances, respectively (see details in Methods and Supplementary
Information). The fitted binding energy curves intersect at around 130.24 G.
Below this magnetic field, the energy of the $AB$ molecule is higher than that
of the AC molecule, and thus the reaction is exoergic. The released energy is
the difference of the molecule binding energies (see Fig. \ref{fig1}\textbf{b})
and is distributed between the products according to energy and
momentum conservation. Therefore, when the kinetic energy acquired by the
reaction products is smaller than the trap depth, the products can be trapped,
which corresponds to the window of $129.7 - 130.2$ G. This narrow window may
be slightly extended towards smaller magnetic fields, because the atom products
can quickly lose their kinetic energy by elastic collisions with background
atoms and thus may still be trapped.

\begin{figure}[ptb]
\includegraphics[width=7cm]{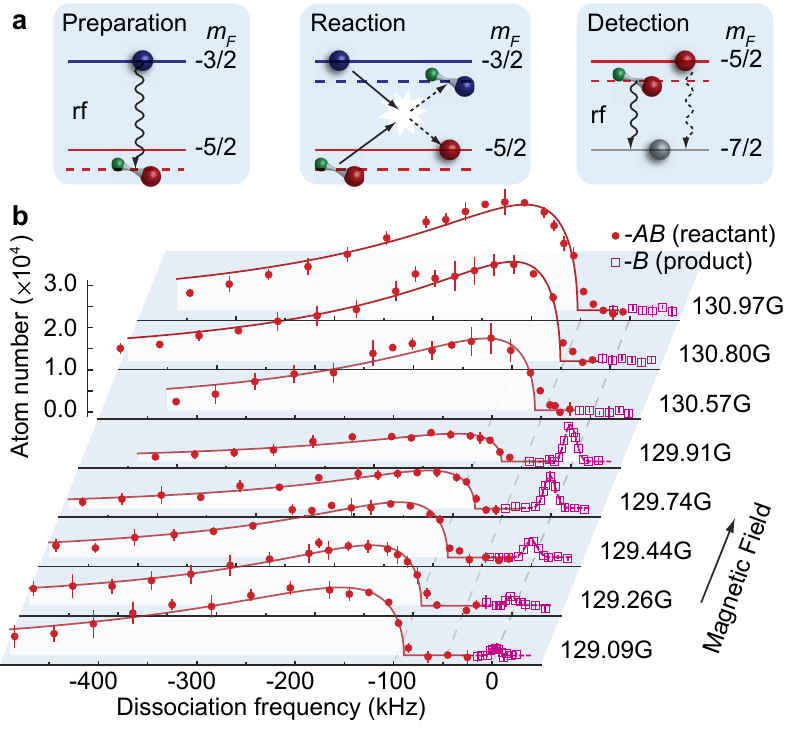}\caption{\textbf{The preparation
and detection scheme and the observation of the atom products. a,} The $AB$
molecules are associated from the $A+C$ mixture by a Blackman rf pulse to
prepare the $AB+C$ mixture ($A$
atoms not shown). After the exchange reaction, the $AB$ molecule and
the $B$ atoms are transferred to the $|9/2, -7/2>$ state for detection by rf
pulses with different frequencies. \textbf{b,} The molecule dissociation
and atom rf spectra. Clear atom peaks are observed for magnetic
fields below 130 G, which implies that the exchange reaction may take place at
these fields. The duration of the association pulse is 0.5 ms (1 ms) for magnetic
fields above (below) 130 G. The $AB$ molecules are dissociated by a 1 ms
square rf pulse applied $50~\mu$s after the association pulse. The atomic rf
spectra are measured by a $57~\mu$s rf pulse, applied $500~\mu$s after the
association pulse. This short pulse is a $\pi$ pulse with an efficiency of
about $90\%$ if on resonance. The dissociation spectra are fitted with the
bound-free Franck-Condon lineshape \cite{Chin2005} and the atomic spectra are
fitted with the Gaussian model. Error bars represent $\pm1$ standard
deviations.}%
\label{fig2}%
\end{figure}

To observe the atom-exchange reaction, we first prepare the atom-dimer mixture
by employing radiofrequency (rf) association \cite{Ospelkaus2006,Zirbel2008,%
Klempt2008,Wu2012}. To this end, as shown in
Fig. \ref{fig2}\textbf{a}, we prepare the Na and K atoms in the $A$ and $C$ states
respectively, then associate the $AB$ molecules from the $A+C$ mixture by
applying a $0.5-1$ ms Blackman rf pulse with a frequency close to the
$|9/2,-3/2\rangle\rightarrow|9/2,-5/2\rangle$ transition (Supplementary
Information). After association, the desired atom-dimer mixture $AB+C$ is
prepared coexisting with the remaining $A$ atoms. If the reaction takes place,
$B$ atoms will appear in the mixture. However, direct absorption imaging
cannot distinguish between the $B$ atoms and $AB$ molecules, and thus we
employ rf spectroscopy to distinguish between them. The $AB$ molecules are
measured by dissociating them into the free atom states $A+|9/2,-7/2\rangle$
for detection. The $B$ atoms are transferred to the $|9/2,-7/2\rangle$ state
for detection by a rf $\pi$ pulse.

\begin{figure*}[htpb]
\includegraphics[width=0.8\textwidth]{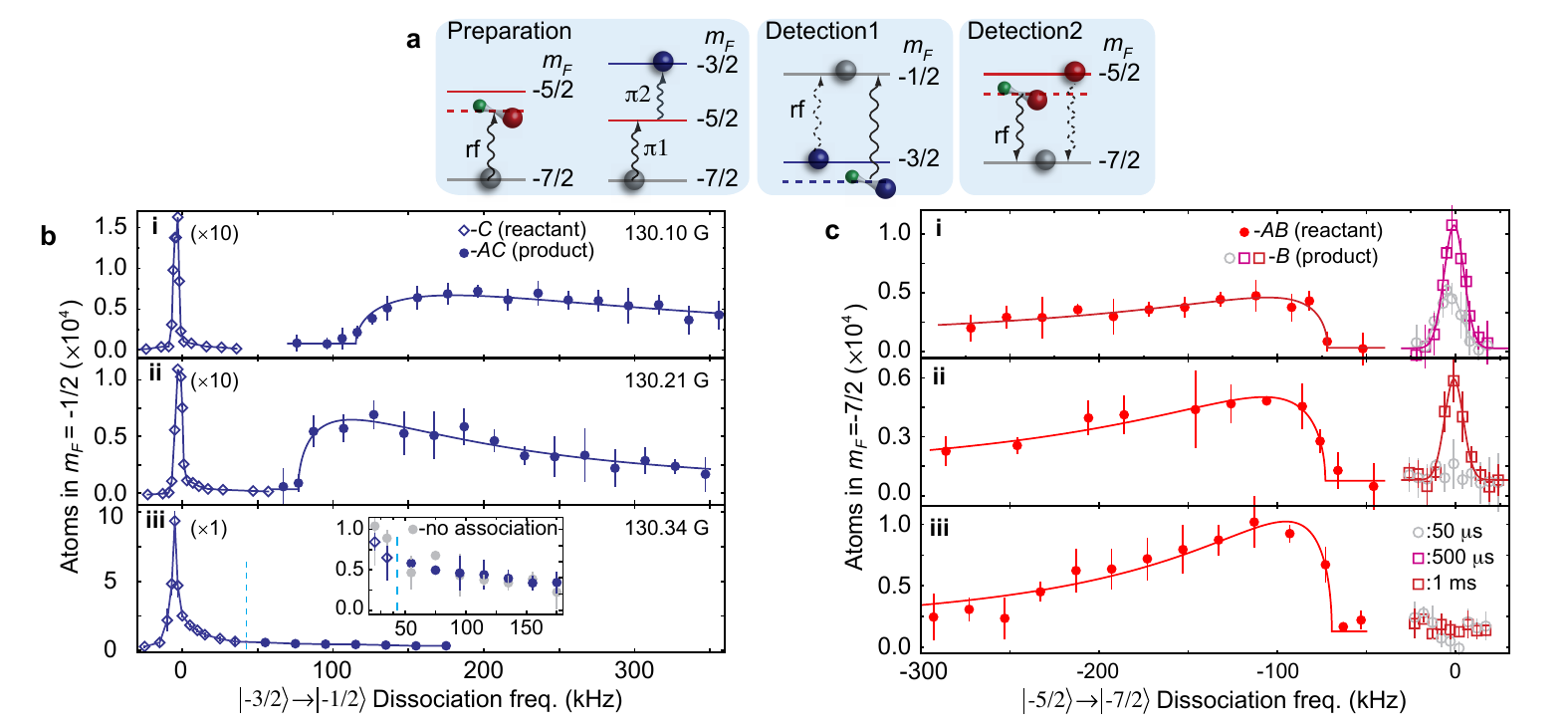}\caption{\textbf{The preparation
and detection scheme and the observation of the molecule and atom products. a,}
The mixture is initially prepared in the $A + |9/2, -7/2\rangle$ state ($A$
atoms not shown), and then the $AB$ molecules are associated with a 1 ms
Blackman pulse, after which the K atoms are transferred into the $C$ state via
two $\pi$ pulses. After the exchange reaction, the $AC$ ($AB$) molecules and
the $C$ ($B$) atoms are transferred into the $|9/2, -1/2\rangle$ ($|9/2,
-7/2\rangle$) state by rf pulses with different frequencies. \textbf{b,} The dissociation
spectra for $AC$ molecules at different magnetic fields. The molecule dissociation
and atom spectra are both measured with 1 ms Blackman rf pulses here, and are
applied $100~\mu$s after the second $\pi$ pulse. Clear $AC$
molecule signals, well separated from the $C$ atom peak, appear at 130.10 G
and 130.21 G. At 130.34 G, no molecule signal is observed, where the gray
points denote the background that no association pulse is applied. The dashed
line represents the molecule binding energy at this field. \textbf{c,} The
atom rf spectra for $B$ atoms. The rf pulse for atoms is a $57~\mu$s short
pulse and is applied $50~\mu$s or $0.5-1$ ms after the second $\pi$ pulse.
Atom increase is clearly observed at 130.10 G and 130.21 G for a $0.5-1$ ms
delay. No atom increase is observed at 130.34 G. Error bars represent
$\pm1$ standard deviations.}%
\label{fig3}%
\end{figure*}

The molecule dissociation and atom rf spectra at different magnetic fields are
shown in Fig. \ref{fig2}\textbf{b}, where the atom and molecule signals are
well separated from each other. Pronounced atom peaks appear at magnetic
fields below 130 G, where the reaction is expected to take place. For magnetic
fields at which the $AB$ molecule has a smaller binding energy, e.g., 130.97 G,
no obvious atomic peaks can be observed. This implies that the observed atom
peaks are not caused by thermal collisional induced dissociation. Note that for
magnetic fields between $130.0$ G and $130.4$ G, rf association of the $AB$
molecules from the $A+C$ mixture is inefficient since the Franck-Condon
coefficient between the free and bound states is largely suppressed when the
scattering lengths of the two channels are almost the same \cite{Chin2005}.

The observation of an atom peak in the magnetic field window may be
considered as an evidence for the atom-exchange reaction. However, it cannot
exclude the possibility that it may be created by simply breaking the $AB$
molecules due to some unclear mechanism. Therefore, an unambiguous
demonstration of the reaction requires the observation of the molecule product
$AC$. Besides, the detection of the molecule product is equivalent to the
direct observation of a spin flip of the Feshbach molecule due to the
spin-exchange interaction.

Detecting the molecule product is much more challenging than detecting the
atom product. In our experiment, there are essentially two difficulties. The first one is
the short lifetime of the molecule product. The second one is that detection of
the $AC$ molecule requires the rf dissociation, which only works well when being
very close to resonance, as this Feshbach resonance is closed-channel
dominated \cite{Chin2005}. Therefore, we choose to work between 130.1 G and
130.4 G. However, association of the $AB$ molecules from the $A+C$ mixture is
inefficient at these fields. Therefore, we change to a different preparation
sequence, as shown in Fig. \ref{fig3}\textbf{a}. We first associate the $AB$
molecules from the $A+|9/2,-7/2\rangle$ mixture, and then apply two successive
$\pi$ pulses to transfer the $|9/2,-7/2\rangle$ atoms to the $C$ state. In
this way the $AB+C$ mixture is also prepared. To observe the $AC$ molecule
products, we dissociate them into the $A+|9/2,-1/2\rangle$ free atom states
for imaging. An advantage of this sequence is that the $|9/2,-1/2\rangle$
state is initially unoccupied and thus even weak dissociation signals can be
well resolved without large backgrounds. To minimize the $AC$ molecule losses
due to inelastic collisions, we switch off the optical trap immediately after
the $\pi$ pulses. The measured spectra are shown in Fig. \ref{fig3}\textbf{b}.
The $AC$ molecule signals, well separated from the atom peaks of the $C$ reactant,
appear at 130.10 G and 130.21 G in the dissociation spectrum, while at 130.34
G no molecule products can be observed. This agrees with the binding energy
measurement that the reaction is endoergic at 130.34 G and thus is forbidden at such a low temperature. Besides, we also measure the atom product
$B$ under this preparation sequence (see in Fig. \ref{fig3}\textbf{c}). To
distinguish the atom products from the background atoms due to the
imperfection of the $\pi$ pulses, we compare the atomic rf spectrum between
two delay times. At 130.10 G and 130.21 G the atom increase can be clearly
observed for the $0.5-1$ ms delay, which confirms that the reaction takes
place. While at 130.34 G no atom increase is observed for 1 ms delay,
which also implies that the reaction is forbidden. The observation of the
molecule product is a "smoking gun" and thus unambiguously demonstrates that
the atom-exchange reaction has been observed.

\begin{figure}[ptb]
\includegraphics[width=7cm]{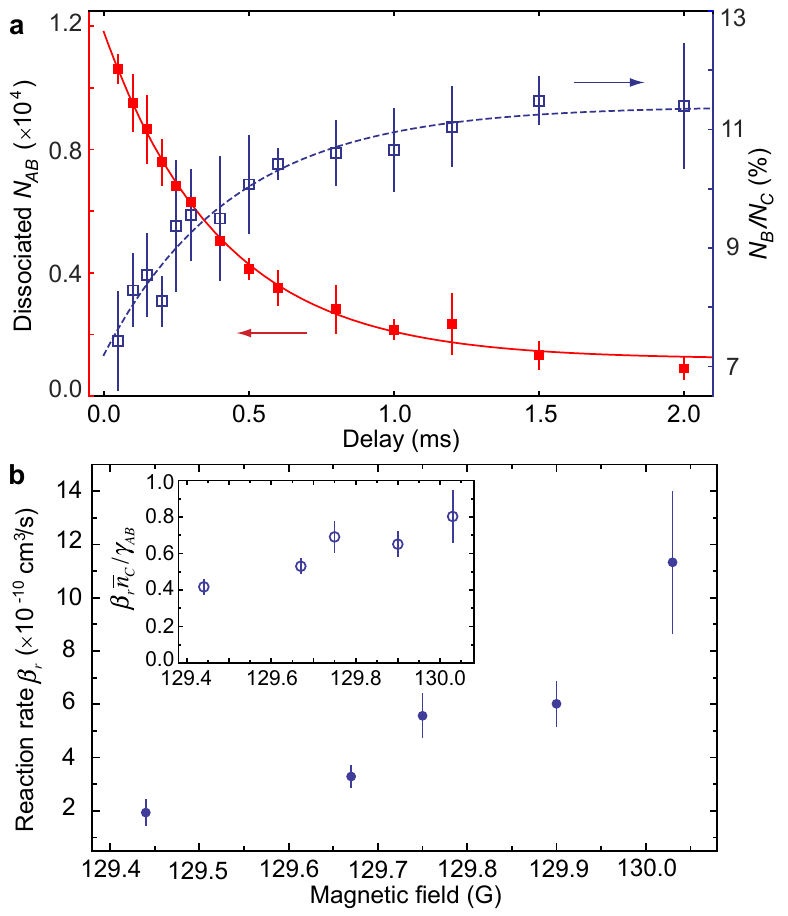}\caption{\textbf{Reaction
dynamics and reaction rate. a,} The time evolution of the ratio between the
number of atom product $N_{B}$ and atom reactant $N_{C}$, and the number of
molecule reactant $N_{AB}$ at 129.90 G. To reduce association time, we use a
0.3 ms square rf pulse to associate the $AB$ molecules. For the atom ratio,
the fitted $1/e$ saturation time is 0.44(8) ms, and the increased ratio is
$4.2(3)\%$. For the $AB$ number, the fitted initial dissociated molecule number  is
$1.06(2)\times10^{4}$, and the $1/e$ lifetime is $0.40(2)$ ms. \textbf{b,} The
measured state-to-state reaction rates at different magnetic fields. (Inset) The
measured ratio between the exchange reaction rate $\beta_{r}\bar{n}_{C}$ and
the total loss rate $\gamma_{AB}$, which shows that the reaction is the dominant
mechanism for molecule losses especially for small released energies. For the
measurement at 130.03 G, we increase the duration of the association rf pulse
to 0.5 ms because the Franck-Condon coefficient is significantly smaller at
this field. Error bars represent $\pm1$ standard deviations.}%
\label{fig4}%
\end{figure}

The observation of the reaction products and the characterization of their
quantum states provide a unique access to studying the reaction dynamics and
measuring the state-to-state reaction rates, which can be achieved by measuring
the time evolutions of the $AB$ reactant and $B$ product numbers. To this
end, we choose to prepare the $AB+C$ mixture with the sequence shown in Fig.
\ref{fig2}\textbf{a}. After a controllable delay, we dissociate the $AB$
molecules to measure the decay of the molecule reactant or apply a $\pi$ pulse
to transfer the $B$ atoms to the $|9/2,-7/2\rangle$ state to measure the
increase of the atom product. The measured time evolution of the reactant
and the product at $129.90$ G is shown in Fig. \ref{fig4}\textbf{a}. It is
readily seen that the reaction is very fast with a time scale of a few hundred
microseconds, during which the losses of the atom reactant $C$ and product $B$
are negligible. The losses of molecule reactants $AB$ due to collisions with
the surrounding atoms cannot be neglected. The reaction dynamics may be
described by
\begin{equation}%
\begin{split}
\dot{N}_{AB}  &  =-\gamma_{AB}N_{AB},\\
\dot{N}_{B}  &  =\beta_{r}\bar{n}_{C}N_{AB},
\end{split}
\label{eqn1}%
\end{equation}
where $N_{AB}$ and $N_{B}$ are the $AB$ reactant and $B$ product numbers,
respectively, $\bar{n}_{C}$ is the mean density of the $C$ reactant,
$\gamma_{AB}$ is the overall decay rate of the $AB$ reactant due to reactive
and inelastic collisions with the surrounding atoms, and $\beta_{r}$ is the
atom-exchange reaction rate to be measured. The time evolution of the atom
product $B$ is exclusively determined by the desired reaction. The reaction
rate can thus be obtained from the initial molecule number $N_{AB}(0)$, the
decay rate $\gamma_{AB}$, and the increase of the ratio $N_{B}/N_{C}$ (see
Methods). The measured state-to-state reaction rates are shown in Fig.
\ref{fig4}\textbf{b}. The reaction rate can be magnetically controlled and
varies from $1.9(3)\times10^{-10}$ cm$^{3}/$s to $1.1(3)\times%
10^{-9}$ cm$^{3}/$s for magnetic fields from 129.44 G to 130.03 G. The
exchange reaction is the dominant molecule loss mechanism. At
$130.03$ G, the measured ratio between the reaction rate $\beta_{r}\bar{n}_C$ and the total decay
rate $\gamma_{AB}$ is $80(14)\%$. This indicates that the reaction selectively reacts
through a single channel, as predicted in Refs.
\cite{Stwalley1978,Stwalley2004}. This dominance may be understood by the
quantum halo character of the Feshbach molecules \cite{Knoop2010}.

The atom-exchange reaction observed in our experiment is in the universal
regime where the interaction between atoms are determined by the large
scattering length \cite{Braaten2006,Incao2009}. In this case, the reaction is
governed by $s$-wave quantum scattering which may be described by the coupled
Skorniakov--Ter-Martirosian (STM) equation \cite{Braaten2009}. There are
unknown three-body parameters which may be determined from the measured
reaction rates. Once the three-body parameters are determined, the scattering
in the whole universal region can be calculated by numerically solving the STM
equation \cite{Braaten2009}.

The observed atom-exchange reaction is also an effective spin-exchange
interaction between the NaK Feshbach molecules and the K atoms. The molecule
and atom products can be considered as the result of spin flips of the molecule
and atom reactants. We expect that at magnetic fields between 130.24 and 130.64 G the
inverse exchange collision $AC+B\rightarrow AB+C$ may also be observed. Such
an effective spin-exchange interaction is just the interaction required to
explore the Kondo effect with NaK Feshbach molecules proposed recently in Ref.
\cite{Bauer2013}, where the NaK Feshbach molecules serve as magnetic
impurities that have spin-exchange interactions with the immersed Fermi sea.
Our work therefore opens up a realistic avenue of quantum simulations of the
Kondo effect and the related strongly correlated phenomenon \cite{Bauer2013}
with an ultracold atom-dimer mixture.

Our work represents the observation of a quantum-state selected/resolved single-channel dominated
ultracold reaction. The quantum states, the temperature, released energy and
the reaction rates are all controllable. One unique feature of this reaction
is that not only the reactants are ultracold, but also the products are in the
ultracold regime since the released energy can be tuned to be very small. This indicates that,
by further lowering the temperature of the K atoms, chemical reactions in quantum
degenerate gases might be studied. For example, if the $B$ state is initially
occupied with a large Fermi energy, and if the released energy in the reaction is
precisely tuned smaller than that Fermi energy, the reaction may be suppressed
due to Pauli-blocking. This opens up the possibility of the investigation of
Pauli-blocked chemistry \cite{Carr2009} which is the fermionic counterpart of
Bose-enhanced superchemistry \cite{Moore2002}.

This work was supported by the National Natural Science Foundation of China,
National Fundamental Research Program of China (Grant No. 2011CB921300), the
Chinese Academy of Sciences.

\section*{Methods}

\textbf{Ultracold Bose-Fermi mixture preparation}. We load Na from a Zeeman
slower and K from a 2D magnetic-optical-trap (MOT) into a two-species
dark-SPOT. The atoms are optically pumped to the $|2,2\rangle$ and
$|9/2,9/2\rangle$ states and are then transferred to an Ioffe-Pritchard cloverleaf
magnetic trap, where Na atoms are subject to forced evaporative cooling and K
atoms are sympathetically cooled. The atomic mixture is then loaded into a
crossed-beam optical dipole trap (wavelength $1064$ nm, beam waist $61$ $\mu$m
and $123$ $\mu$m for horizontal and vertical beams respectively) and Na atoms
are transferred to the $A$ state before further evaporative cooling. At
the end of the optical trap evaporation we adiabatically increase the optical
trap power to the desired value to hold the atoms and perform the experiments.
The Na atoms are always in the $A$ state, and the K atoms are prepared in
different internal states by a $50$ ms adiabatic rapid passage rf sweep with
nearly unit efficiency. The experiments are performed at magnetic fields of
around 130G. The magnetic field is actively stabilized with a stability of
better than 10 mG. We prepare K atoms in the $|9/2,-1/2\rangle$ or
$|9/2,-7/2\rangle$ states at a low field before increasing the magnetic field to
the desired value, since these two states have no Feshbach resonances at about
130 G. K atoms in the $C$ state are prepared from the
$|9/2,-1/2\rangle$ state by a $\pi$ pulse transfer at high field. K atoms in
different internal states are measured with a high-field imaging
technique, and the measured atoms numbers are calibrated by comparing with the
$\sigma-$ cycling transition of $|9/2,-9/2\rangle$.

\textbf{Characterization of the Feshbach resonances}. The binding energy of the
$AB$ Feshbach molecule is measured by the rf loss or dissociation
spectrum. For the rf loss spectrum, we prepare the K atoms in the $C$ state,
and apply a $0.5-1$ s rf pulse to couple the free atom states to the
molecular bound states, where the rf field has a Rabi frequency of $\sim1$ kHz
for the free atomic transition. The binding energy is then obtained by fitting
the loss spectrum (Supplementary Information) with models in Ref.
\cite{Ulmanis2015} . For the dissociation spectrum, we associate the $AB$
molecules from the $A+C$ mixture and then dissociate them into the
$A+|9/2,-7/2\rangle$ state as in the main text. The binding energy is obtained
by fitting the dissociation spectrum using the bound-free Franck-Condon
lineshape in Refs. \cite{Chin2005,Wu2012}.

The binding energy of the $AC$ Feshbach molecule is measured using similar
methods. For the rf loss spectrum, we prepare free K atoms in the $|9/2,-1/2\rangle$
state, and apply a weak rf pulse to observe the loss spectrum. For the dissociation
spectrum, the $AC$ molecules are rf associated from the $A+|9/2,-1/2\rangle$ state.
Then the remaining K atoms in the $|9/2,-1/2\rangle$ state are transferred to the
$|9/2,1/2\rangle$ state via a $\pi$ pulse, leaving the $|9/2,-1/2\rangle$
state empty for the molecule dissociation. The $AC$ molecules has to be
dissociated in such a way because the dissociation from the $AC$ molecule into the
$A+B$ free state is significantly suppressed due to the existence of the
bound-bound transition \cite{Chin2005}.

We first fit the measured binding energies with the universal model
\cite{chin2010}, $E_{b}=-\hbar^{2}/2\mu(a-\bar{a})^{2}$, where $\bar
{a}=51~a_{0}$ is the mean scattering length with $a_{0}$ the Bohr radius, and
$a=a_{\mathrm{bg}}[1-\Delta B/(B-B_{0})]$ is the scattering length near the
Feshbach resonance with the background scattering length $a_{\mathrm{bg}},$
the resonance position $B_{0}$ and the width $\Delta B$ the fitting parameters.
For Feshbach resonance between $A$ and $B,$ the fitting yields $a_{\mathrm{bg}%
}= - 455(18)~a_{0}$, $B_{0}=138.71(20)$ G and $\Delta B=-34.60(34)$ G, which
are in good agreement with previous work \cite{Wu2012}. This is an open
channel dominated resonance with strength $s_{\mathrm{res}}\gg1$ \cite{Viel2016}.
For Feshbach resonance between $A$ and $C,$ we obtain
$a_{\mathrm{bg}}=126(9)~a_{0}$, $B_{0}=130.637(14)$ G and $\Delta
B=4.0(4)$ G. This resonance tends towards closed channel dominance with the
strength $s_{\mathrm{res}}<1$ \cite{Viel2016}. The binding energy measurement
gives a resonance at 130.64 G instead of 129.4 G determined from the enhanced atom
loss measurement \cite{Park2012}. The measured binding energies have also been
fitted with the coupled-channel calculations (Supplementary Information).

\textbf{Reaction dynamics}. The time evolution of the molecule reactant number
$N_{AB}$ and atom product number $N_{B}$ may be described by Eqn.
(\ref{eqn1}), where the overall loss rate is $\gamma_{AB}=\beta
_{A}\bar{n}_{A}+\beta_{C}\bar{n}_{C}+\beta_{r}\bar{n}_{C}$, with $\bar{n}_{A}$
and $\bar{n}_{C}$ mean densities of the $A$ and $C$ atoms, respectively. Here
the first term is the loss due to inelastic collisions with the remaining $A$
atoms, with $\beta_{A}$ the loss rate. The other two terms describe the losses
due to collisions with the atom reactant $C$, which include the desired
reactive collision with reaction rate $\beta_{r}$, and the losses due to
reactions in other channels and inelastic collisions with a loss rate of
$\beta_{C}$. The collisions between the $AB$ molecules can be safely neglected
since the Feshbach molecules are fermionic molecules. As the number of molecule
reactants is about one order smaller than that of the atoms, we assume $\bar
{n}_{A}$ and $\bar{n}_{C}$ are constants during the reactions.

Eqns. (\ref{eqn1}) can be solved with the solution,
\begin{align}
N_{AB}(t) &  =N_{AB}(0)e^{-\gamma_{AB}t},\\
N_{B}(t) &  =\frac{\beta_{r}\bar{n}_{C}N_{AB}(0)}{\gamma_{AB}}(1-e^{-\gamma
_{AB}t})+N_{B}(0),
\end{align}
where $N_{B}(0)$ describes the atom product number accumulated during
the association process of the $AB$ molecules. Therefore, the decay rate of the
reaction may be given by
\begin{equation}
\beta_{r}=\frac{\Delta N_{B}\gamma_{AB}}{\alpha N_{C}N_{AB}(0)}%
\end{equation}
where $\Delta N_{B}=N_{B}(\infty)-N_{B}(0)$ is the increased atom product
number after the reactions, the mean density of the atom reactant is $\bar{n}_{C}=%
\alpha N_{C}$, with $\alpha=(\frac{m_{\mathrm{K}}\bar{\omega}^2}{4\pi k_{B}%
T_{\mathrm{K}}})^{3/2}$, $\bar{\omega}$ the geometric mean of the trapping
frequencies of the K atoms and $T_{\mathrm{K}}$ the temperature of the K
atoms. Therefore the reaction rate may be obtained by measuring the increased
ratio between the atom product and reactant $\Delta N_{B}/N_{C}$, the initial
molecule reactant number $N_{AB}(0)$, and the decay rate of the $AB$ molecule
$\gamma_{AB}$. Note that the reaction is very fast, part of the $AB$ molecules may
have been reactively lost during the dissociation process. Thus the rf dissociation
rate also needs to be taken into account to correct the initial
molecule number (Supplementary Information). Statistical and systematical
uncertainties in the atom number, molecule number, molecule loss rate and
temperature have all been included to calculate the uncertainty in the final
reaction rate.

\newpage

\section*{Supplementary information}

\subsection{Coupled channel calculation}

In the main text, the measured binding energy is fitted using the simple
universal model. Here we perform the coupled-channel calculation and compare
the theory with the experimental results. The Hamiltonian describing the $s$-wave
scattering is
\begin{equation}
\renewcommand\theequation{S\arabic{equation}} H=T+\sum_{S=0,1}V_{S}%
(r)P_{S}+H_{hf}+H_{z}.
\end{equation}
The first term is the kinetic energy $T=-\frac{\hbar^{2}}{2\mu}\frac{d^{2}}{ dr^{2}}$
with $\mu$ the reduced mass. The second term describes the spin-exchanging
interaction, where $P_{0}=1/4-\mathbf{s}_{\alpha}\cdot\mathbf{s}_{\beta}$ and
$P_{1}=3/4+\mathbf{s}_{\alpha}\cdot\mathbf{s}_{\beta}$ are the singlet and
triplet projection operator respectively with $\mathbf{s}$ the electron spin.
Here and below we refer to Na as $\alpha$ and K as $\beta$. $V_{0}(r)$ and
$V_{1}(r)$ denotes the Born-Oppenheimer singlet potential $X^{1}\Sigma$ and
triplet potential $a^{3}\Sigma$. The Born-Oppenheimer potentials can be
expressed as power expansion of $r$, whose latest version can be found in Ref.
\cite{Temelkov2015s}. $H_{hf}$ is the hyperfine interaction term, described by%
\begin{equation}
\renewcommand\theequation{S\arabic{equation}} H_{hf}=a_{hf\alpha}%
\mathbf{s}_{\alpha}\cdot\mathbf{i}_{\alpha}+a_{hf\beta}\mathbf{s}_{\beta}%
\cdot\mathbf{i}_{\beta},
\end{equation}
where $a_{hf}$ is the hyperfine constant and $\mathbf{i}$ is the nuclear spin.
The last term is the Zeeman term
\begin{equation}
\renewcommand\theequation{S\arabic{equation}} H_{z}=[(g_{s\alpha}s_{z\alpha
}-g_{i\alpha}i_{z\alpha})+(g_{s\beta}s_{z\beta}-g_{i\beta}i_{z\beta})]\mu
_{B}B_z,
\end{equation}
with $g_{s}$ the electron g-factor, $g_{i}$ the nuclear g-factor and $B_{z}$
the bias magnetic field.

The internal state may be expressed in terms of the spin basis $|\sigma
\rangle=|m_{i_{\alpha}},m_{s_{\alpha}};m_{i_{\beta}},m_{s_{\beta}}\rangle.$
The Hamiltonian couples all the internal states with the same $M_{F}%
=m_{f_{\alpha}}+m_{f_{\beta}}$ with $m_{f}=m_{i}+m_{s}$. For a given $M_{F}$
and $B_{z}$, we first diagonalize the $H_{hf}+H_{z}$ to obtain the internal
eigenstate $|\chi_{i}\rangle$ and the threshold energy $E_{i}^{th}$ of each
channel. Expanding the wave function in terms of the new bases $|\psi
\rangle=\sum_{i}\psi_{i}(r)|\chi_{i}\rangle$, we obtain the coupled channel
Schr\"{o}dinger equation
\begin{equation}
\renewcommand\theequation{S\arabic{equation}} \sum_{j}[T\delta_{ij}%
+\sum_{S=0,1}V_{S}(r)\langle\chi_{i}|P_{S}|\chi_{j}\rangle]\psi_{j}%
(r)=(E-E_{i}^{th})\psi_{i}(r)
\end{equation}
for a given entrance energy $E$. The scattering length and the binding energy
are calculated using standard multichannel log-derivative method. In our
numerical calculations, we integrate from $1${\AA } up to $5000${\AA } with a
step of $10^{-3}${\AA }.

There are in total three Feshbach resonances between $^{23}$Na $|1,1\rangle$ and
$^{40}$K $|9/2,-5/2\rangle$ ($|9/2,-3/2\rangle)$) for a magnetic field of up to 300 G.
Therefore, the coupled-channel calculated scattering length is fitted using the formula
$a=a_{\mathrm{bg}}(1+\eta B)(1-\frac{\Delta B_{1}}{B-B_{1}})(1-\frac{\Delta B_{2}}{B-B_{2}%
})(1-\frac{\Delta B_{3}}{B-B_{3}})$~\cite{Jachymski2013s,Viel2016s}, where
$\eta$ is a small parameter taking into account the slow variation of the
background scattering length as a function of the magnetic field. The results are
shown in Table \ref{tab.feshbachres}.

\begin{table}[hbp]
\renewcommand\thetable{S\arabic{table}}
\begin{tabular}{|c|c|c|}
\hline
& $|1,1\rangle\otimes|9/2,-5/2\rangle$ & $|1,1\rangle\otimes|9/2,-3/2\rangle$ \\
\hline
$a_{\mathrm{bg}}(a_{0})$ & -488 & -442\\
\hline
$\eta$ & -0.001 & -0.001\\
\hline
$B_{1}$(G) & 96.64 & 117.27\\
\hline
$B_{2}$(G) & 107.09 & 130.78\\
\hline
$B_{3}$(G) & 137.14 & 177.63\\
\hline
$\Delta B_{1}$(G) & 2.40 & -2.10\\
\hline
$\Delta B_{2}$(G) & 3.83 & 4.83\\
\hline
$\Delta B_{3}$(G) & -41.42 & -57.64\\
\hline
\end{tabular}
\caption{Feshbach resonance parameters given by the coupled-channel calculations. }
\label{tab.feshbachres}
\end{table}

\begin{figure}[ptb]
\includegraphics[width=0.9\columnwidth]{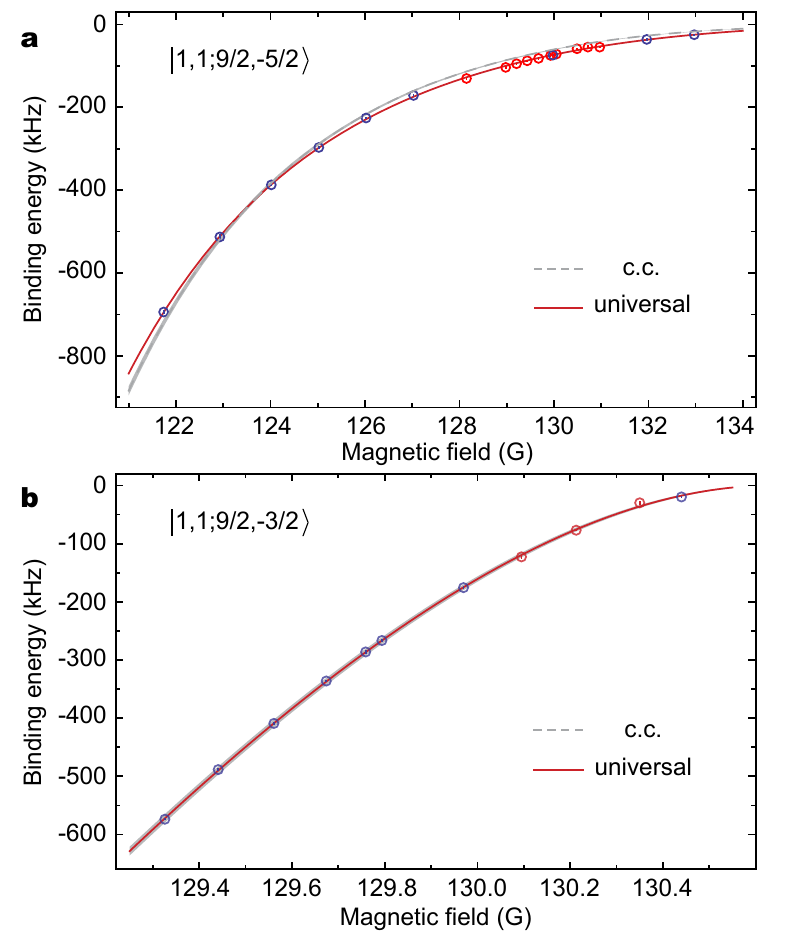}
\renewcommand\thefigure{S\arabic{figure}}
\caption{The comparison between the coupled-channel calculations and the experimental results.
The red (blue) open circle represents the binding energy measured from the rf dissociation
(loss) spectrum. The gray shaded region corresponds to the uncertainty of the binding
energy fitted with the coupled-channel calculations. All error bars represent $\pm1$
standard deviations of the statistical uncertainties. }
\label{fig_feshbachres}%
\end{figure}

It can be readily seen that the Feshbach resonance positions are slightly
different from the experimental results. Therefore we fit the measured binding
energies by using the coupled-channel calculations with the resonance position as
the only fitting parameter \cite{Zirbel2008s}. The results are shown in
Fig.~\ref{fig_feshbachres}.

For the Feshbach resonance between $|1,1\rangle$ and $|9/2,-5/2\rangle$, the universal
model gives a resonance position of $B_{\mathrm{0}}=138.71(20)$ G. The fitting using the
coupled-channel calculations gives a resonance position of $B_{\mathrm{0}}=137.38(5)$ G.
It can be seen in Fig.~\ref{fig_feshbachres} that the c.c. result does not agree very well
with the experimental data. This may imply that the parameters of the potential are still
not accurate enough for this open-channel dominated resonance. For the Feshbach
resonance between $|1,1\rangle$ and $|9/2,-3/2\rangle$, the universal model gives a
resonance position of $B_{\mathrm{0}}=130.637(14)$ G. The fitting using the coupled-channel
calculations gives a resonance position of $B_{\mathrm{0}}=130.635(1)$ G. For this
resonance, the experiment results and the coupled-channel calculations agree very well
with each other. Note that a systematical uncertainty of 10 mG in the magnetic field
is not included in the above analysis.


\subsection{RF loss spectrum}

To characterize the Feshbach resonances, we measure the binding energy of the
Feshbach molecules either using the rf dissociation spectrum or the rf loss
spectrum. The dissociation spectrum has been discussed in the main context.
Here we explain the details of the rf loss spectrum.

In our experiment, we use a weak and long square rf pulse to couple the
free atomic state to the bound molecular state and to observe the
atom losses with respect to the rf frequency as in Ref.~\cite{Ulmanis2015s}. As
the associated molecules are lost quickly by inelastic collisions with the
surrounding atoms (dominately with Na atoms), one molecule loss roughly
corresponds to the loss of one K atom and two Na atoms. We measure the total
atom number of $N_{\mathrm{K}}+0.5N_{\mathrm{Na}}$, which roughly corresponds
to twice of the associated molecule numbers. Besides, we find empirically that
the total atom number has a best stability when adding the K atom number with
one half to one fourth of the Na atom number. The observed atom loss spectrum
is then fitted with the model introduced in Ref.~\cite{Ulmanis2015s,Klempt2008s}
for each magnetic field, which includes the density of the relative motion of
the atom pair, the bound-free Franck-Condon factor, and a collisional
broadening profile.

\begin{figure}[tpbh]
\includegraphics[width=0.9\columnwidth]{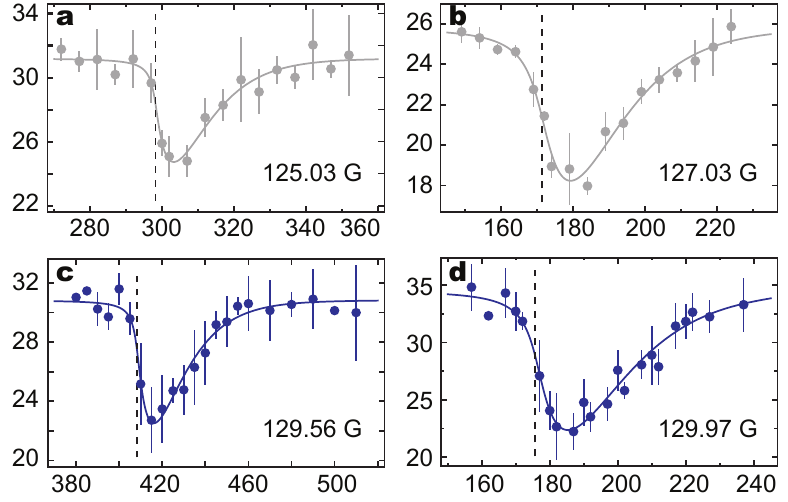}
\renewcommand\thefigure{S\arabic{figure}}
\caption{Loss spectrum of the Na and K atoms due to the rf coupling of the free-%
bound transition. \textbf{a} and \textbf{b}, the free scattering state $A+C$ is coupled
to the bound $AB$ dimer state. The duration of the rf pulse is 0.5 s for both
measurements. \textbf{c} and \textbf{d}, the free scattering state $A+|9/2,-1/2\rangle$
is coupled to the bound $AC$ dimer state. The duration of the rf pulse is 1.0 s for
129.97 G and 1.2 s for 129.56 G. For the horizontal axis, the free atomic transition
frequency of K is defined to be zero. For the vertical axis, it represents the total atom
number of $N_{\mathrm{K}}+0.5N_{\mathrm{Na}}$ with unit of $10^4$. The dashed
line represents the position of the fitted dimer binding energy. The association rf pulse
has a Rabi frequency of about $2\pi\times1$ kHz for the free atomic transition. The
mixture temperature is about 600 nK for these measurements. All error bars represent
$\pm1$ standard deviations. }\label{fig_lossspec}%
\end{figure}

The measured atom loss spectra at several magnetic fields are shown in
Fig.~\ref{fig_lossspec}. In \textbf{a} and \textbf{b}, the atom mixture is
prepared in the $A+C$ state, and it's coupled to the bound $AB$ dimer state
with a weak association pulse. The fitting gives a binding energy of $171.4(9)$
kHz and $297.0(7)$ kHz for the magnetic field of 127.03 G and 125.03 G,
respectively. In \textbf{c} and \textbf{d}, the atom mixture is prepared in the
$A+|9/2,-1/2\rangle$ state, it's coupled to the bound $AC$ dimer state with
the weak association pulse. The fitting gives a binding energy of $175.6(6)$
kHz and $409.1(12)$ kHz for the magnetic field of 129.97 G and 129.56 G,
respectively.


\subsection{Association spectrum of the Feshbach molecule}

\begin{figure}[ptbh]
\includegraphics[width=0.9\columnwidth]{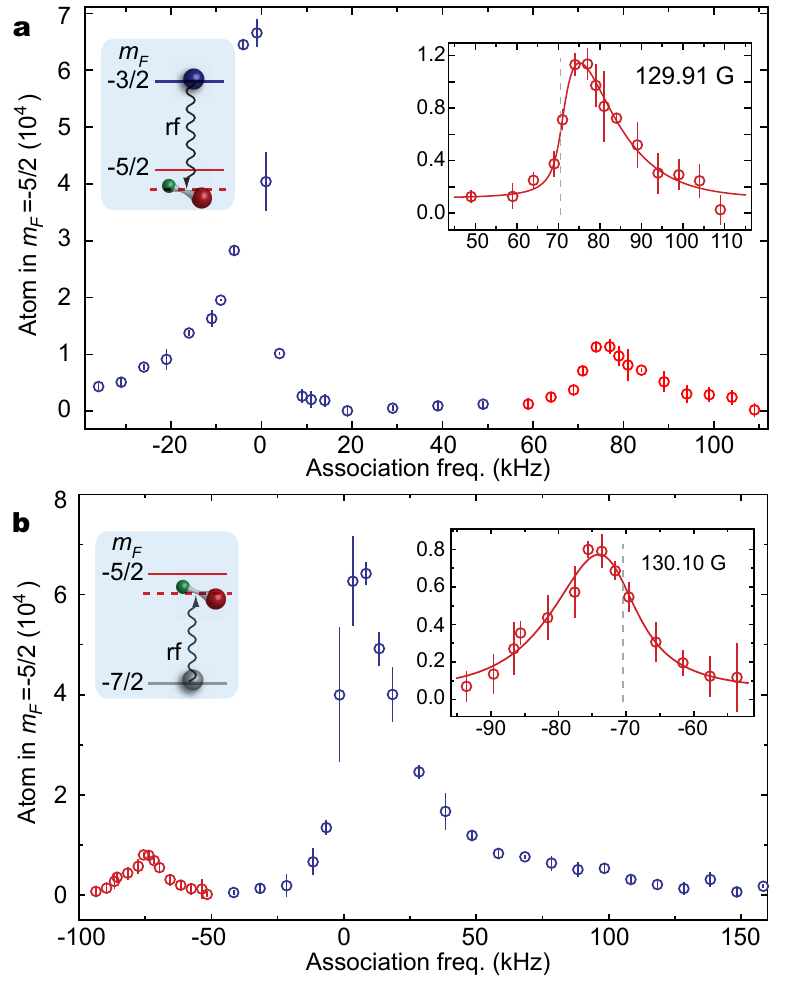}
\renewcommand\thefigure{S\arabic{figure}}
\caption{Association spectrum of the $AB$ Feshbach molecule from the $A+C$
or $A+|9/2,-7/2\rangle$ mixture. The free atomic transition frequency is set to be
zero. And the long tails of the free K atomic transition is due to the mean field
shift from the interaction with the overlapping Na atoms. (Inlet) Molecule
association spectrum (free Na atoms are not shown). The curves are fitted with
the model in Ref.~\cite{Ulmanis2015s,Klempt2008s}. The dashed lines represent
the position of the fitted binding energy of the molecules. All error bars represent
$\pm1$ standard deviations. }\label{fig_assospec}%
\end{figure}

In our experiment, we associate the Feshbach molecules by applying a rf pulse
to transfer the free scattering atom pairs ($A+C$ or $A+|9/2,-7/2\rangle$) to the
$AB$ molecule. For different measurements, the association rf field can be either
Blackman or square pulses, with a peak Rabi frequency of about $2\pi\times20$
kHz for the free atomic transition.

For the observation of the atom product $B$ at different magnetic fields as in
Fig.~2 of the main text, we use the Blackman rf pulses for molecule association
to narrow the spectral width of the free atomic transition (suppress the
Fourier side lobes compared with square pulses). Typical association spectra
with the Blackman rf pulse are shown in Fig.~\ref{fig_assospec}, which is
taken by directly imaging the K atoms in the $B$ state after association. Note
that the imaging itself does not distinguish between the free $B$ atoms and the $AB$
molecules. It's obvious that the atom peak and molecule peak are well
separated in the rf spectrum. This means that at the molecule association
frequency, the background $B$ atoms, directly transferred from the $C$ state
via free atomic transition, is negligible. The is important since it implies
that the observed $B$ atoms in Fig.~2 are not directly transferred by the
association rf pulse, but created from the chemical reactions.

For the measurement of the reaction dynamics, we use square rf pulse to
associate the $AB$ Feshbach molecules from the $A+C$ mixture. The use of
square wave pulse can reduce the length of the association pulse. This square rf
pulse will transfer at most about 5\% of the $C$ atoms to the $B$ state for
detuning of larger than 4 times the Rabi frequency. This can only contribute a
constant background in the atom product signal. The increase of the atom
product with time can only be caused by reaction, which is used to measure the
reaction rate.


\subsection{Time scale of the dissociation procress}

In studying the reaction dynamics, the Feshbach molecules are dissociated into
free atoms for detection by a rf pulse with a duration of about 1 ms and a
peak Rabi frequency of about $2\pi\times16$ kHz for the free atomic
transition. As discussed in the main text, the time scale of the reaction is
about several hundred microseconds. Therefore, in the dissociation process, part of the molecules are lost
due to reaction and thus cannot be dissociated into atoms for detection.

We may derive the relation between the dissociated molecule number and the
initial total molecule number as follows. When applying the dissociation
pulse, the time evolution of the molecule may be described by,

\begin{equation}
\renewcommand\theequation{S\arabic{equation}}
\begin{split}
\dot{N}_{\mathrm{mol}} &= -\gamma_{\mathrm{loss}}N_{\mathrm{mol}} - \gamma_{\mathrm{diss}}N_{\mathrm{mol}}, \\
\dot{N}_{\mathrm{atom}} &= \gamma_{\mathrm{diss}} N_{\mathrm{mol}},
\end{split}
\end{equation}
where $N_{\mathrm{mol}}$ is molecule number and $N_{\mathrm{atom}}$ is the
atom number obtained by dissociation, and $\gamma_{\mathrm{loss}}$ describes
the losses of the molecules via inelastic
and reactive collisions with other atoms (inverse of the molecule lifetime),
and $\gamma_{\mathrm{diss}}$ is the dissociation rate. As the background atom
number is about one order larger than the molecule number, we assume a
constant loss rate of $\gamma_{\mathrm{loss}}$. This equation is
straightforward to solve with the solution,
\begin{equation}
\renewcommand\theequation{S\arabic{equation}} N_{\mathrm{atom}}(t_{\mathrm{rf}%
})=\frac{\gamma_{\mathrm{diss}}N_{\mathrm{mol}}(t_{\mathrm{rf}}=0)}{\gamma_{\mathrm{diss}%
}+\gamma_{\mathrm{loss}}}(1-e^{-\gamma_{\mathrm{diss}}t_{\mathrm{rf}}%
-\gamma_{\mathrm{loss}}t_{\mathrm{rf}}}),
\end{equation}
where $t_{\mathrm{rf}}$ is the duration of the applied dissociation rf pulse. For a
sufficiently long dissociation rf pulse, we have
\begin{equation}
\renewcommand\theequation{S\arabic{equation}} N_{\mathrm{atom}}(t_{\mathrm{rf}}=\infty
)=\frac{\gamma_{\mathrm{diss}}}{\gamma_{\mathrm{diss}}+\gamma_{\mathrm{loss}}%
}N_{\mathrm{mol}}(t_{\mathrm{rf}}=0).\label{eqn.atommolnum}%
\end{equation}
Here $N_{\mathrm{atom}}(t_{\mathrm{rf}}=\infty)$ is just the dissociated molecule
number in the dissociation process. Thus we can obtain the initial number of molecules
from the measured dissociated molecule number.

\begin{figure}[ptb]
\includegraphics[width=0.9\columnwidth]{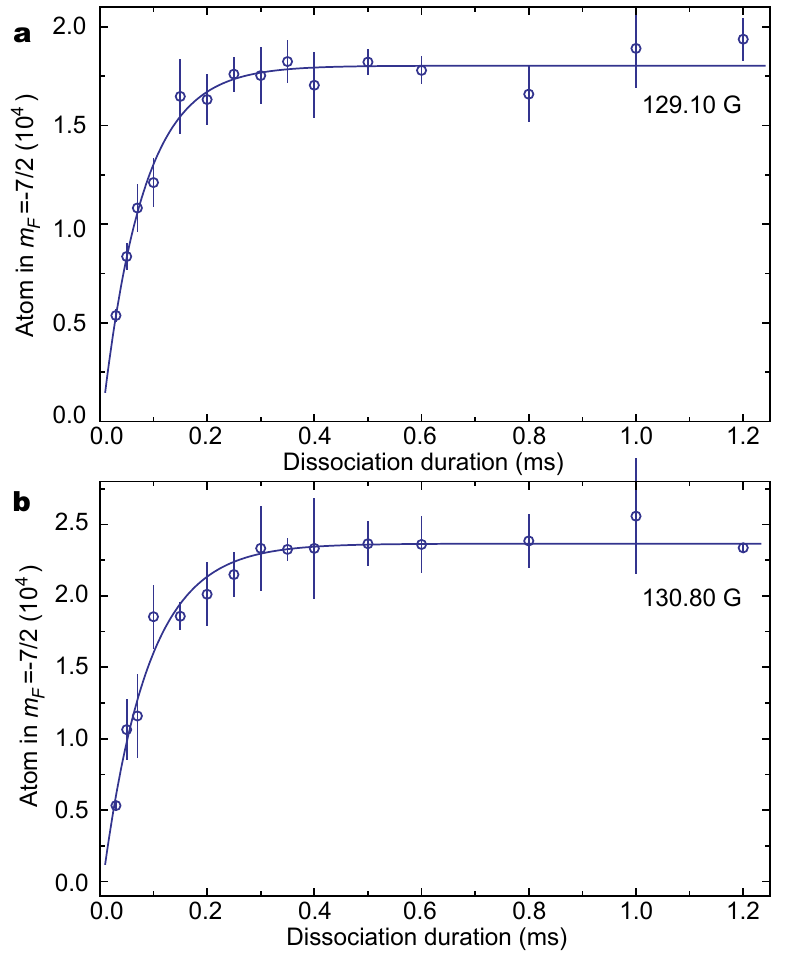}
\renewcommand\thefigure{S\arabic{figure}}
\caption{Saturation of the rf dissociation of Feshbach molecules. The measurements
are taken with square rf pulses with frequencies at $50\sim100$ kHz below the
bound-free transition (onset) frequency. The lines represent the fitted results with
the model of $N_{\mathrm{atom}}(t_{\mathrm{rf}})=N_{0}(1-e^{-\gamma_{\mathrm{diss}}%
t_{\mathrm{rf}}})$ where we have neglected $\gamma_{\mathrm{loss}}$ for simplicity.
All error bars represent $\pm1$ standard deviations. }\label{fig_dissduration}%
\end{figure}


The dissociation rate $\gamma_{\mathrm{diss}}$ is estimated as follows. We
associate the $AB$ Feshbach molecules from the $A+C$ mixture and then use a
square rf pulse with variable durations to dissociate the $AB$ Feshbach
molecules into the $A+|9/2,-7/2\rangle$ state. Then we measure the number of K
atoms in the $|9/2,-7/2\rangle$ state as a function of rf pulse duration. This measurement
is carried out at the magnetic fields of 129.10 G and 130.80 G where the
lifetime of molecule is larger than 1.4 ms. The results are shown in
Fig.~\ref{fig_dissduration}.

For the magnetic field of 129.10 G, the dissociation frequency is 150 kHz
below the frequency of the free atomic transition of $B\rightarrow
|9/2,-7/2\rangle$, while the energy of the $AB$ molecule is about $h\times96$ kHz
below that of the free $B$ state. The dissociation rate of this dissociation
process is fitted to be $\gamma_{\mathrm{diss}}=12.56(82)$ kHz, where we have
neglected $\gamma_{\mathrm{loss}}$ since it is about one order smaller than
$\gamma_{\mathrm{diss}}$. For the magnetic field of 130.80 G, the dissociation
frequency is 130 kHz below the free atomic transition frequency, while the
energy of the $AB$ molecule is about $h\times56$ kHz below the free $B$ state. The
saturation rate of the dissociation process is measured to be $\gamma
_{\mathrm{diss}}=10.93(80)$ kHz. In the reaction dynamics measurement, we
assume a constant dissociation rate and use a mean value of $\bar{\gamma
}_{\mathrm{diss}}=11.7(8)$ kHz to correct the initial molecule numbers from
the measured dissociated molecule number.


\subsection{Reaction dynamics measurement}

\label{sec.reactiondynamic}

\begin{figure}[ptbh]
\includegraphics[width=0.9\columnwidth]{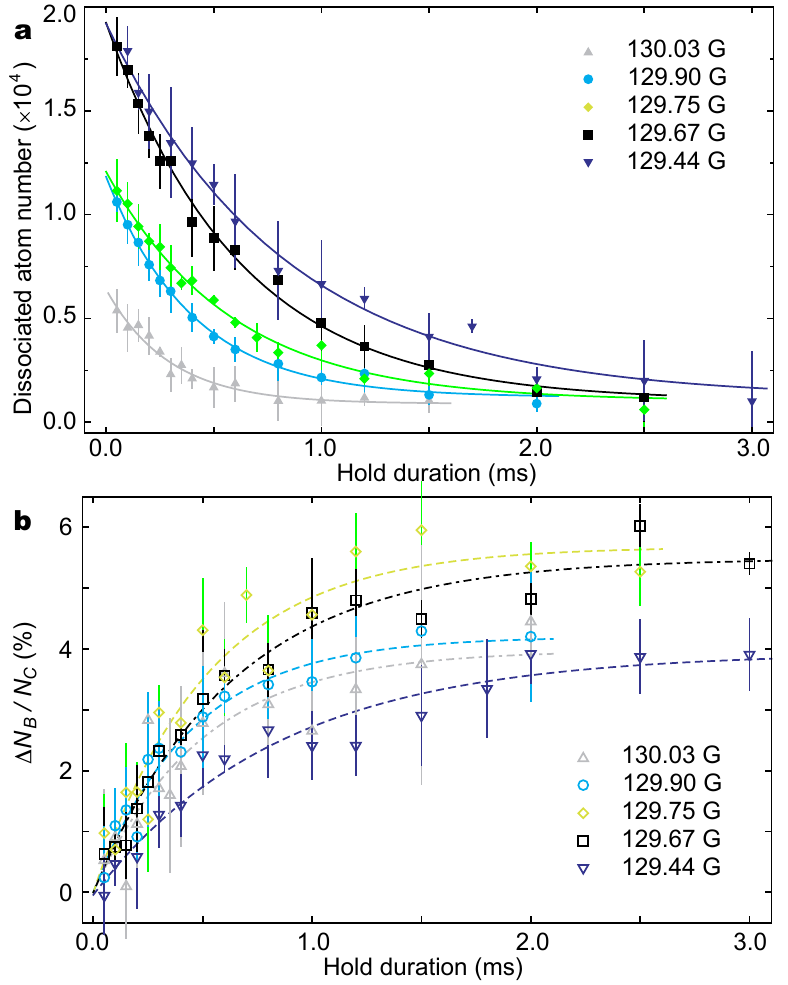}
\renewcommand\thefigure{S\arabic{figure}}
\caption{Time evolution of the number of $AB$ reactant and $B$ product. All
error bars represent $\pm1$ standard deviations.}\label{fig_reacdym}%
\end{figure}

In the main text, we have explained how to derive the exchange-reaction rate
from the measurement of time evolutions of the molecule reactant number and
the atom product number. The measured time evolution at 129.90 G is given as
an example for simplicity. Here we provide all measured results for the
reaction rates given in Fig. 4 of the main text. The reaction rate can be
given by,
\begin{equation}
\renewcommand\theequation{S\arabic{equation}} \beta_{r}=\frac{\Delta N_{B}%
}{\alpha N_{C}} \frac{\gamma_{AB}}{N_{AB}(0)},
\end{equation}
where $\alpha=(\frac{m_{\mathrm{K}} \bar{\omega}^{2}}{4\pi k_{\mathrm{B}}
T_{\mathrm{K}} })^{3/2}$ is a density coefficient which gives the mean atom
density in the $C$ state as $\bar{n}_{C}=\alpha\times N_{C}$. The temperature
of the $K$ atoms in the $C$ state is measured after molecule rf association at
each magnetic field, which typically gives a temperature of about 650 nK.
Together with the measured trapping frequency for the K atoms in the optical
trap, we can get the value of this density coefficient.

To measure the time evolution of the number of the molecule reactant $AB$, we
dissociate them into the free atom pairs of $A+|9/2,-7/2\rangle$ and measure
the K atom number in the $|9/2,-7/2\rangle$ state, after holding the $AB+C$
mixture in the optical trap for a specific duration. Then the time evolution of the
dissociated molecule number (K atoms in the $|9/2,-7/2\rangle$ state) is fitted
with an exponential decay model as shown in Fig.~\ref{fig_reacdym}\textbf{a},
which gives the dissociated molecule number $N^{\rm{diss}}_{AB}(0)$ for zero
holding duration and the molecule lifetime $1/\gamma_{AB}$. Then the initial
molecule number $N_{AB}(0)$ can be obtained from Eqn.~S\ref{eqn.atommolnum},
where the loss rate $\gamma_{\mathrm{loss}}$ is equivalent to the overall loss
rate of $\gamma_{AB}$.

For the measurement of the product atoms in the $B$ state, we use a rf $\pi$
pulse to transfer the $B$ atoms to the $|9/2,-7/2\rangle$ state for absorption
imaging. The measured time evolution of the atom number ratio is then fitted
with an exponential saturation model, $A\times(1-\exp{(-t/\tau)})$, as shown
in Fig.~\ref{fig_reacdym}\textbf{b}. Note that only the increased part
the $B$ atoms is needed for analysis here. Besides, $\Delta N_{B}$ is the increased
atom number measured in the $|9/2,-7/2\rangle$ state, which is transferred
from the $B$ state with the $\pi$ pulse. Thus a finite transfer efficiency of
the $\pi$ pulses, which is about $\eta_{\pi}=90\%$, has to be taken into
account to give the correct increased ratio.

Finally, the reaction rate is given by
\begin{equation}
\renewcommand\theequation{S\arabic{equation}}
\beta_{r}=\frac{\Delta N_{B}}{\eta_{\pi}\alpha N_{C}}\frac{\gamma_{AB}%
\gamma_{\mathrm{diss}}}{(\gamma_{AB}+\gamma_{\mathrm{diss}})N^{\mathrm{diss}}_{AB}(0)}.
\end{equation}
The reaction rates given in the main text are calculated in this way. All
statistical and systematical uncertainties of the parameters in the above
equation are included to give the uncertainty of the calculated reaction rates.

\end{document}